\documentclass[runningheads]{llncs}

\usepackage[frozencache]{minted}
\usepackage{ffcode}
\usepackage{mdframed}
\usepackage{longtable}
\usepackage{lscape}
\usepackage{href-ul}
\usepackage{mathtools}

\begin{document}

\title{Detecting unanticipated mutual recursion using Elegant Objects representation of object-oriented programs}
\titlerunning{Detecting unanticipated mutual recursion using EO}

\author{
%  Mansur Khaziev \hspace{0.5em}
   Nikolai Kudasov \hspace{0.5em}
   Mikhail Olokin \hspace{0.5em}
   Oleksii Potyomkin \hspace{0.5em} \\
   Nikolay Shilov \hspace{0.5em}
   Maxim Stepanov}
%\author{
%  Mansur Khaziev \and %\hspace{0.5em}
%  Nikolai Kudasov \and
%  Mikhail Olokin \and
%  Alexey Potyomkin \and \\
%  Nikolay Shilov \and
%  Maxim Stepanov}
\authorrunning{N.~Kudasov, M.~Olokin, O.~Potyomkin, N.~Shilov, and M.~Stepanov}
\institute{Innopolis University, Innopolis, Russia
\email{\{n.kudasov,a.potyomckin,n.shilov\}@innopolis.ru\\ \{m.stepanov, m.olokin\}@innopolis.university}}
\maketitle
%\author{Nickolay Shilov}
%\email{n.shilov@innopolis.ru}
%\author{Violetta Sim}
%\email{v.sim@innopolis.ru}
%\author{Maxim Stepanov}
% \affiliation{%
%   \institution{Innopolis University}
%   \streetaddress{Universitetskaya 1}
%   \city{Innopolis}
%   \state{Tatarstan Republic}
%   \postcode{420500}
%   \country{Russia}
% }

% \renewcommand{\shortauthors}{Kudasov, Shilov, and Stepanov}

\newcommand\todo[1]{\textcolor{red}{TODO: #1}}
\newcommand{\java}[1]{\mintinline{java}{#1}}
\newcommand{\cpp}[1]{\mintinline{cpp}{#1}}
\newcommand{\python}[1]{\mintinline{python}{#1}}
\newcommand{\eo}[1]{\ff{#1}}

%%

\begin{abstract}
Elegant Objects (EO) is a variation of the object-oriented programming paradigm that favors pure objects and decoration.
EO programming language is based on these ideas and has been suggested by Bugayenko as an intermediate representation for object-oriented programs.
This paper provides plausible representations in EO of some class-based constructions from Java, C++, and Python. We then reformulate the classical fragile base class problem in the context of these representations. Finally, we discuss an algorithm for detecting a subset of fragile base class patterns in EO programs. We show that using EO as an intermediate language is plausible and discuss possible improvements to the language to assist in richer static analysis.
\end{abstract}





% \maketitle

\section{Introduction}

Conventional object-oriented languages provide flexible tools for software developers. However, these languages do not fully isolate the implementation of a class from its subclasses. The lack of proper protection makes it hard to refactor or modify base classes, as such modifications may lead to breakage in subclass code.

\emph{Fragile base class problem} is a category of problems involving classes that cannot always be safely modified without also updating subclasses correspondingly. The two main culprits in many fragile base class problems are \emph{open recursion} and \emph{method overriding}.

\begin{figure}
  \begin{minted}{java,firstnumber=last,linenos,numbersep=2pt,framesep=6pt,framerule=1pt,highlightcolor=gray!30,rulecolor=gray,frame=leftline,escapeinside=||,mathescape,style=bw}
  class A {
    int x;
    void f(int y) { this.x = y; }
    void g(int y) { f(y); }
  }

  class B extends A {
    @Override
    void f(int z) { g(z); }
  }
  \end{minted}
  \caption{Unanticipated mutual recursion in class \java{B}.}
  \label{figure:java-unanticipated-recursion}
\end{figure}

In the Java code snippet shown in Figure~\ref{figure:java-unanticipated-recursion}, class \java{A} has two methods, both of which perform the same action. Class \java{B} inherits from \java{A} and overrides the definition of method \java{f} to reuse the code in \java{g}. However, since method \java{g} is defined in a way that involves calling method \java{f} of the object, calling method \java{f} of an instance of class \java{B} will result in an infinite recursion.

A concern here is not with the recursion itself but with how it is introduced. Inlining the call \java{f()} yields
\begin{minted}{java,firstnumber=last,linenos,numbersep=2pt,framesep=6pt,framerule=1pt,highlightcolor=gray!30,rulecolor=gray,frame=leftline,escapeinside=||,mathescape,style=bw}
void g(int y) { this.x = y; }
\end{minted}
Such a change does not change the semantics of class \java{A}: each object of this class behaves the same before and after inlining. 
However, in this example, the inlining of \java{f()} changes the semantics of class \java{B}~--- now, there is no recursion.

Fragile base class problem is a concern for software engineers, especially for library designers, since they do not control the user code where subclasses are defined.
To mitigate this problem, many developers advocate for avoiding inheritance. Instead, Bloch \cite{Bloch2018}, Szyperski \cite{Szyperski2002}, and others suggest using delegation by wrapping base class instance in its original state and explicitly forwarding control when necessary. Bloch \cite{Bloch2018} also recommends that library designers make their classes \java{final} to disable the possibility of inheritance altogether.
%\debate[Yegor]{There is one more way to solve this problem: make the base class ``final.'' This is what PMD, for example, will recommend you do if you pass your Java code through its analysis. Maybe it's worth mentioning this.}

% TODO: existing approaches to solve fragile base class: avoid inheritance, use delegation, document

% TODO: existing approaches to fragile base class detection: ?

% somewhat contradicting ideas: delegation used instead of inheritance, while decoration is the only feature for object extension. (inheritance ~ object extension) + delegation is not an Elegant Objects paradigm's idea, it's rather a workaround for having dynamic self  
Bugayenko \cite{bugayenko2021eolang} introduced the Elegant Objects paradigm, taking the idea of using delegation instead of inheritance to the absolute. Moreover, the EO programming language provides decoration as the only language feature for object extension. As we present in Section~\ref{section:fragile}, using decorators makes it easier to recognize when an EO program relies on open recursion. This is instrumental for the detection algorithm in Section~\ref{section:detecting-mutual-recursion}.
% Maybe worth mentioning that EO relying on open recursion is not that bad (because it's lazy and declarative) — contrary to mainstream object oriented languages (they are strict and open recursion is the thing to be avoided)

In this paper, we propose using the EO programming language as an intermediate representation for the analysis of object-oriented programs. EO is a minimalistic programming language and more convenient in formalizations than full-featured languages such as Java or C++. Using EO for intermediate representation also allows preserving the important structural properties of the original object-oriented code. We present an approach for detecting specific fragile base class problems in object-oriented programs based on this representation.

Our paper is structured as follows:
\begin{enumerate}
  \item Section~\ref{section:translation} presents an approach for translating class-based constructions from Java, C++, and Python to the EO programming language. We show that such a translation is faithful concerning open recursion and pre- and post-conditions of object methods.
  \item In section~\ref{section:fragile}, we revisit the classical fragile base class problem and reformulate it as a ``fragile decorated objects'' problem for EO. We show that the translation in Section~\ref{section:translation} preserves the problem. We emphasize a specific case of the problem~--- unanticipated mutual recursion.
  \item section~\ref{section:detecting-mutual-recursion} provides an algorithm for detecting unanticipated mutual recursion in EO programs.
  \item Section~\ref{section:implementation} explains the implementation of the algorithm in Scala, discussing the steps involved in analyzing the EO code.
  \item Section~\ref{section:benchmarking-methodology} discusses a methodology for benchmarking static analysis tools and describe input/output formats and used metrics.
  \item In section~\ref{section:benchmarking-results}, we discuss the results of benchmarking Polystat and Clang-Tidy, giving an interpretation of the report.
  \item In section~\ref{section:discussion}, we discuss the benefits and limitations of EO as an intermediate language for the analysis of object-oriented programs. We suggest possible modifications of EO that might make it a more attractive option for static analysis.
\end{enumerate}

\section{Translating  classes to Elegant Objects}
\label{section:translation}
%\debate[Shilov]{Emulating Classes in EO}
% TODO: FIX LABELS (they are currently reused; that breaks references)

This section describes a possible translation of class-based constructions from well-known object-oriented languages to EO programming language. EO is an object-oriented language, but it does not have a notion of a class. So, we translate classes into objects capable of producing new objects~--- instances of the said class.

Versions used at the time of writing this paper:
\begin{itemize}
    \item EO: 0.21.0
    \item Objectionary: 0.21.0
\end{itemize}

We present a simplified translation aimed at keeping the structural properties of the source code useful for further analysis. We leave the details of memory management, types, static class methods, interfaces, input/output, and execution mechanisms out of the scope of this paper.

EO is a minimalistic homoiconic language, similar to languages like Self, Lisp, and Prolog, and any expression in EO is an object. An essential feature of EO is that the only way to extend objects is via decoration. With decoration, we can construct a new object with another object (decoratee) to ask for attributes that the enclosing object does not have.

%\begin{example}
  In the following code, object \eo{y} decorates object \eo{x}. Then, object \eo{z} decorates \eo{y}:
\begin{minted}{text,firstnumber=last,linenos,numbersep=2pt,framesep=6pt,framerule=1pt,highlightcolor=gray!30,rulecolor=gray,frame=leftline,escapeinside=||,mathescape,style=bw}
[] > x
  1 > a
  2 > b
[] > y
  x > @
  3 > a
[] > z
  y > @
  4 > c
\end{minted}
  Evaluating \eo{z.a}, \eo{z.b}, and \eo{z.c} would yield 3, 2, and 4 correspondingly.
%\end{example}

%\debate[Yegor]{Your use of Examples is a bit inconsistent. Here, for example, you present this code as Example 2.1, but never mention this in the text. You just expect the reader to read this example in their nature flow of reading the rest of the text. In other places you call them Figures and ask the reader to "look at the Fig. 1". I think some consistency is needed here.}

% Different languages provide different approaches to implement classes. EO does not have any classes in itself, so the choice of such is up to the developer.

%\subsection{Mutable object state in EO}

%The ultimate goal of converting classes from traditional OOP languages is to get rid of mutability, as EO is meant to be immutable. This poses a problem, because all of modern OOP languages rely heavily on mutable state. In the current state, EO provides workarounds to simplify conversion of existing code to EO, such as \texttt{memory} and \texttt{cage} for storing primitive types (numbers, strings, chars and byte arrays) and objects accordingly, in a mutable attribute.

% TODO: write about approaches to remove mutability: SSA, heap/serializers (implemented in EO).

\subsection{Classes as object factories}

We translate classes to the so-called \emph{object factories}~--- objects that can produce new objects. In particular, each object factory will have the attribute \eo{new} to produce a new instance of the corresponding class. The simplest such class is \eo{classObject}, which we will use as a base class for all other classes:
\begin{minted}{text,firstnumber=last,linenos,numbersep=2pt,framesep=6pt,framerule=1pt,highlightcolor=gray!30,rulecolor=gray,frame=leftline,escapeinside=||,mathescape,style=bw}
[] > classObject
  [] > new
\end{minted}

Attributes declared in a class definition translate directly to attributes of an object constructed with \eo{new}. To facilitate initialization, we use \eo{seq} construction and \eo{memory} in EO. The translation of a Java class definition is presented in line~no.~\ref{example:java-class-to-eo}.

%\begin{example}
  Consider the following class definition in Java:
  \begin{minted}{java,firstnumber=last,linenos,numbersep=2pt,framesep=6pt,framerule=1pt,highlightcolor=gray!30,rulecolor=gray,frame=leftline,escapeinside=||,mathescape,style=bw}
class A {
  int i = 0;
  A(int i) { this.i = i; }
}
  \end{minted}

The following EO code corresponds to the Java code above (description is provided after the snippet):
\begin{minted}{text,firstnumber=last,linenos,numbersep=2pt,framesep=6pt,framerule=1pt,highlightcolor=gray!30,rulecolor=gray,frame=leftline,escapeinside=||,mathescape,style=bw}
[] > classA |$\label{example:java-class-to-eo}$|
  classObject > @
  [] > new
    classObject.new > super
    [] > this
      super > @
      memory > i
      [this i] > run_constructor
        seq > @
          this.i.write i
          this
    seq > @
      this.i.write 0
      this
  [j] > constructor
    new > this
    seq > @
      this.run_constructor this j
      this
\end{minted}
%\label{example:java-class-to-eo}
%\end{example}

%\debate[Yegor]{I think this code won't compile. For example, at the line \eo{classObject.new new} --- what is the second "new"? Also, who will ever call "constuctor" and why? Besides, what's the point of showing double underscrore lines in this small paper? I understand that it's important for the actual software, but in the paper we can simplify the code for the sake of readability and call classes like "classA," for example.}

We note that here, we split the construction of a new object into several parts:
\begin{enumerate}
  \item First, \eo{classA.new} initializes an object with proper attributes and initial values.
  \item The initialized object has a method \eo{run\_constructor} that runs the code corresponding to the constructor defined in the original Java class.
  \item Then, \eo{classA.constructor} combines two methods to initialize an object and run the constructor code on it.
\end{enumerate}

With this presentation, object initialization in Java
\begin{minted}{java,firstnumber=last,linenos,numbersep=2pt,framesep=6pt,framerule=1pt,highlightcolor=gray!30,rulecolor=gray,frame=leftline,escapeinside=||,mathescape,style=bw}
A a = new A(3);
\end{minted}
will correspond to the following EO program:
\begin{minted}{text,firstnumber=last,linenos,numbersep=2pt,framesep=6pt,framerule=1pt,highlightcolor=gray!30,rulecolor=gray,frame=leftline,escapeinside=||,mathescape,style=bw}
classA.constructor 3 > a
\end{minted}

Class methods are translated with an explicit \eo{this} argument in the first position, and all object method calls are modified to pass the object itself as \eo{this} in every method call:

%\debate[Yegor]{Starting the caption of a figure with "Consider..." is wrong, I believe. You should say what it is, not what you want the reader to do. If you want to go with the "Consider.." format (which I think is the best), I would suggest you get rid of the idea of wrapping your code snippets in "figure" environment. Just let them be part of the text, like you do a few lines later, directly using "minted" environment. Also, I would suggest you using "ffcode" package of mine (it's in CTAN) with "ffcode" environment. It also provides "ff" command which would replace your "eo" one. }

%\begin{example}
  Consider adding two more methods to the class definition in line~no.~\ref{example:java-class-to-eo}:
  \begin{minted}{java,firstnumber=last,linenos,numbersep=2pt,framesep=6pt,framerule=1pt,highlightcolor=gray!30,rulecolor=gray,frame=leftline,escapeinside=||,mathescape,style=bw}
public class A {
  ...
  public void f(int x) { this.i = x; }
  public void g(int y) { f(y + 1); }
}
  \end{minted}

  The corresponding EO code is extended as follows:
  \begin{minted}{text,firstnumber=last,linenos,numbersep=2pt,framesep=6pt,framerule=1pt,highlightcolor=gray!30,rulecolor=gray,frame=leftline,escapeinside=||,mathescape,style=bw}
[] > classA
  classObject > @
  [] > new
      ...
      [this x] > f
        seq > @
          this.i.write x
          this
      [this y] > g
        seq > @
          this.f this (y.add 1)
          this
      ...
  \end{minted}
%\end{example}

We note that static class methods, even though irrelevant in the context of this paper, could be translated similarly, except they would belong to the class object and would not take \eo{this} argument. 

Finally, inheritance is replaced with decoration under the translation to EO. A subclass is represented as an object that decorates the superclass object, and the \eo{new} method explicitly calls the superclass version of \eo{new}:

Consider the following class definition in Java:
  \begin{minted}{java,firstnumber=last,linenos,numbersep=2pt,framesep=6pt,framerule=1pt,highlightcolor=gray!30,rulecolor=gray,frame=leftline,escapeinside=||,mathescape,style=bw}
class B extends A {
  int j;
  B() { super(1); this.j = 3; }
  void f(int x) { this.i = x + this.j; }
}
  \end{minted}

  The corresponding EO code is
  \begin{minted}{text,firstnumber=last,linenos,numbersep=2pt,framesep=6pt,framerule=1pt,highlightcolor=gray!30,rulecolor=gray,frame=leftline,escapeinside=||,mathescape,style=bw}
[] > classB
  classA > @
  [] > new
    classA.new > super
    [] > this
      super > @
      memory > j
      [this] > run_constructor
        seq > @
          super.run_constructor super 1
          this.j.write 3
          this
    this > @
  [] > constructor
    new > this
    seq > @
      this.run_constructor this
      this
  [this x] > f
    seq > @
      this.i.write (x.add this.j)
      this
  \end{minted}

% Here, the word "instead" (`super` instead of `this`) is causing confusion: what is implied? That there was an earlier example were `run_constructor` of super class was called with `this`, and here we call it with `super`)
% Apparently it is not the case, probably what was meant is that we call all methods are called with `this`, and run_constructor of the superclass is an exception to this tacit rule, — which is not entirely obvious to the reader.
Observe that in \eo{run\_constructor} we call the constructor code the superclass using \eo{super.run\_constructor} and pass \eo{super} instead of \eo{this}. It is done to ensure that we initialize correctly, in case some attributes or methods of \eo{classB} shadow attributes or methods of \eo{classA}.

%\debate[Shilov]{translate --> emulate, translation --> emulation?}
As presented here, the translation loses some of the information, such as types and access qualifiers. This loss is acceptable for the purposes of analysis of the fragile base class problem discussed in this paper. However, a more detailed translation mechanism should be possible for wider applications.
% "Wider"? or "broader", or "more comprehensive"?

\subsection{Translating from C++}

Translating from C++ is straightforward, except for memory-related primitives such as dereferencing pointers in the presence of pointer arithmetic. At the time of writing, EO does not provide tooling for proper manual memory management. Still, as we are interested in the hierarchical structure of the code more than the actual execution, we suggest, for analysis, replacing any low-level code that cannot be converted into EO with an expression that forces the evaluation of subexpressions (for example, by printing their values).

  Consider the following class definition in C++:
  \begin{minted}{cpp,firstnumber=last,linenos,numbersep=2pt,framesep=6pt,framerule=1pt,highlightcolor=gray!30,rulecolor=gray,frame=leftline,escapeinside=||,mathescape,style=bw}
class A {
  int i = 0;
  public:
    A(int i) { this->i = i; }
    void f(int x) { this->i = x; }
    void g(int y) { this->f(y+1); }
}
  \end{minted}

%\debate[Yegor]{You can make code more compact by simply adding the section "public:" to it. Also, no need to make the class "public" in this paper --- this is irrelevant to the problem being discussed. Also, why do you prefix variables with "this->" but don't do the same with the call of "f(y+1)"?}

  The following EO code corresponds to the C++ code above:
  \begin{minted}{text,firstnumber=last,linenos,numbersep=2pt,framesep=6pt,framerule=1pt,highlightcolor=gray!30,rulecolor=gray,frame=leftline,escapeinside=||,mathescape,style=bw}
[] > classA
  classObject > @
  [] > new
    classObject.new > super
    [] > this
      super > @
      memory > i
    seq > @
      this.i.write 0
      this
  [j] > constructor
    new > this
    seq > @
      this.i.write j
      this
  [this x] > f
    seq > @
      this.i.write x
      this
  [this y] > g
    seq > @
      this.f this (y.add 1)
      this
  \end{minted}

\subsection{Translating from Python}

Translating Python classes is straightforward, with a significant difference that Python methods are declared with an explicit \python{self} argument, so we do not have to add one when translating declarations.
However, we still have to pass the object as \python{self} for method calls:

%\begin{example}
  Consider the following class definition in Python:
  \begin{minted}{python,firstnumber=last,linenos,numbersep=2pt,framesep=6pt,framerule=1pt,highlightcolor=gray!30,rulecolor=gray,frame=leftline,escapeinside=||,mathescape,style=bw}
class A:
  i = 0
  def init(self, i):
    self.i = i
  def f(self, x):
    self.i = x
  def g(self, y):
    self.f(y + 1)
  \end{minted}

  The following EO code corresponds to the Python code above:
  \begin{minted}{text,firstnumber=last,linenos,numbersep=2pt,framesep=6pt,framerule=1pt,highlightcolor=gray!30,rulecolor=gray,frame=leftline,escapeinside=||,mathescape,style=bw}
[] > classA
  classObject > @
  [] > new
    classObject.new > super
    [] > self
      super > @
      memory > i
      [self i] > run_constructor
        seq > @
          self.i.write i
          self
    seq > @
      self.i.write 0
      self
  [j] > constructor
    new > self
    seq > @
      self.run_constructor self j
      self
  [self x] > f
    seq > @
      self.i.write x
      self
  [self y] > g
    seq > @
      self.f self (y.add 1)
      self
  \end{minted}
%\label{example:python-class-to-eo}
%\end{example}

\section{Fragile decorated objects}
\label{section:fragile}

In this section, we revisit the fragile base class problem and reformulate it as a \emph{fragile decorated object problem} in the context of the EO programming language. In particular, we focus on the unanticipated mutual recursion variant of the problem. This reformulation serves as a foundation for the defect detecting algorithm discussed in Section~\ref{section:detecting-mutual-recursion}.

\subsection{Fragile base class problem}

The fragile base class problem is an anti-pattern in object-oriented code. It defines scenarios where modifications in the base class not affecting its objects' behavior can still affect the behavior of the objects of its subclasses. Mikhailov and Sekerinski \cite{MikhajlovSekerinski1998} have identified and formalized five types of fragile base class. We reproduce their definitions in a specialized context.

In our research, we are concerned with the static analysis of code without considering the changes to the source code of classes. Our setting requires a proper adaption of the fragile base class problem, as we no longer can work with a single specified base class refinement. We have one version of the code for the base class, so we have to assume some possible refinements instead.
%\debate[Yegor]{I'm a bit lot here. What is "evolution of classes"? The rest of the paragraph is also unclear. I didn't understand what you are talking about here.}

To get plausible refinements automatically from a single base class version, we consider inlining method definitions in the call site. Consider the following definition:

\begin{minted}{java,firstnumber=last,linenos,numbersep=2pt,framesep=6pt,framerule=1pt,highlightcolor=gray!30,rulecolor=gray,frame=leftline,escapeinside=||,mathescape,style=bw}
class A {
  int n = 0;
  void f(x) { this.n = x; }
  void g(y) { f(y+1); }
}
\end{minted}

Inlining the definition of method \java{f} at the call site yields:

\begin{minted}{java,firstnumber=last,linenos,numbersep=2pt,framesep=6pt,framerule=1pt,highlightcolor=gray!30,rulecolor=gray,frame=leftline,escapeinside=||,mathescape,style=bw}
class A {
  int n = 0;
  void f(x) { this.n = x; }
  void g(y) { this.n = y+1; }
}
\end{minted}

%\debate[Yegor]{I think this example is a bit redundant. You already mentioned "inlining" at the first page of the paper. You also provided an example there. I suggest explain inlining in more details right there, at the top of the paper and remove this duplicated explanation from here.}

We note that many modern object-oriented programming languages lack referential transparency. Replacing a variable or a method with its value or definition does not always produce an equivalent program. Thus, direct inlining is not always a faithful 
%\debate[Shilov]{a faithful --> an equivalent?}
program transformation. However, in EO, we safely inline methods definitions (attributes, whose value is an abstract object), as in EO, objects are pure and method applications do not force argument evaluation (EO has essentially a version of call-by-need evaluation).
%\debate[Yegor]{Broken paragraph? By the way, what is "referential transparency"? Maybe it would be to explain in a few words?}

%\debate[Shilov]{The following definitions are borrowed from *somewhere* or are original?}
\begin{definition}
  We call class $A'$ a \emph{refinement} of class $A$ if it can be produced from $A$ via inlining none, one, or more method calls.
\end{definition}

\begin{definition}
  A class $A$ is called \emph{evidently fragile} in program $P$ if there exists a refinement $A'$ of $A$ and a descendant class $D$ that inherits, perhaps indirectly, from $A$, such that replacing $A$ with $A'$ changes observational properties of class $D$.
  %TODO: reformulate. Replace refinement with refactoring?
\end{definition}

\begin{definition}
  A class $A$ is called \emph{fragile} if there exists a program $P$ such that $A$ is evidently fragile in $P$.
\end{definition}

\subsubsection*{Unanticipated mutual recursion}

This paper focuses on just one type of fragile base class problem~--- unanticipated mutual recursion in a subclass. Mutual recursion happens when two or more methods recursively call each other. Such recursion is not a problem on its own. However, when such recursion relies crucially on implementation details in a base class, this behavior may be fragile and change under modifications to the base class code. In particular, consider the following example:

\begin{minted}{java,firstnumber=last,linenos,numbersep=2pt,framesep=6pt,framerule=1pt,highlightcolor=gray!30,rulecolor=gray,frame=leftline,escapeinside=||,mathescape,style=bw}
class A {
  void f() { return 3; }
  void g() { return f(); }
}
class B extends A {
  @Override
  void f() { return g(); }
}
\end{minted}

Here, calling method \java{f()} of an object \java{b} of class \java{B} results in calling \java{b.g()}, which in turn calls \java{b.f()} again, leading to infinite mutual recursion. However, if we inline the call to \java{f()} in the implementation of \java{A.g}, the behavior of \java{b.f()} changes drastically~--- it will result in calling \java{b.g()} that would return 3.
%\debate[Yegor]{You've said this already in the beginning of the paper and explained what is the problem. This paragraphs only repeat what we already know. I suggest getting rid of them somehow.}

Although we have infinite mutual recursion in this minimal example, in general, any mutual recursion that is unstable under refinements of a base class is considered unanticipated.

%TODO: when is such an anti-pattern useful?

%\subsubsection*{Unjustified assumptions in subclasses}

%To be done.

\subsection{Reformulating with Elegant Objects}

As the EO programming language does not have classes or inheritance, we reformulate the fragile base class problem in terms of objects and decoration.

\subsubsection*{Decoration and attribute shadowing}

Method overriding is a crucial ingredient in the fragile base class problem. In EO, we use decoration with attribute shadowing instead. Consider the following example:
\begin{minted}{text,firstnumber=last,linenos,numbersep=2pt,framesep=6pt,framerule=1pt,highlightcolor=gray!30,rulecolor=gray,frame=leftline,escapeinside=||,mathescape,style=bw}
[] > a
  1 > x
  2 > y
[] > b
  a > @
  3 > x
\end{minted}

Here, object \python{b} decorates object \python{a}, but has its own attribute \python{x}. So \python{b.x} evaluates to 3, while \python{b.y} evaluates to 2 since it has to come from the decoratee.

Another important detail in the context of classes is that methods implicitly accept and use a reference to the object of a given class. For example, in Java code in Figure~\ref{figure:java-unanticipated-recursion}, method calls \java{f(...)} and \java{g(...)} can be replaced with a more explicit \java{this.f(...)} and \java{this.g(...)} correspondingly. In Python, method definitions explicitly mention parameter \python{self}, even though its value is introduced implicitly on call site. In EO, according to translation presented in Section~\ref{section:translation}, we rely on \eo{self} argument explicitly both in method definition and when calling a method of an object:

\begin{minted}{text,firstnumber=last,linenos,numbersep=2pt,framesep=6pt,framerule=1pt,highlightcolor=gray!30,rulecolor=gray,frame=leftline,escapeinside=||,mathescape,style=bw}
[] > a
  [self x] > f
    x > @
  [self y] > g
    self.f self y > @
\end{minted}

We note that the name and position of the attribute \eo{self} is a convention and is not a special syntax of EO programming language.

\subsubsection*{Refinement of objects}

Similarly to refinements of classes, we can speak about refinements of objects in EO. However, in the presence of methods that accept \eo{self} explicitly, it becomes a little less straightforward. Note that we now want to inline the call \eo{self.f self y}. Since \eo{self} is an argument to \eo{a.g}, we have to make an assumption that \eo{self} is the same as \eo{a} to be able to use the definition of \eo{a.f}.

To adapt the notion of refinement to objects, we introduce several definitions.
First, we specify how we identify the call sites of interest:

\begin{definition}
  An application \eo{s.f a1 a2 ... aN} is called an \emph{inline candidate}
  if \eo{s} is a void attribute\footnote{void attributes in EO are uninstantiated attributes that can be instantiated at most once using object application}
  %\debate[Shilov]{What is "void attribute"?}
  bound by some enclosing abstract object,
  \eo{f} is an attribute identifier,
  and \eo{a1}, \eo{a2}, \ldots \eo{aN} are argument expressions such that
  at least one of the arguments is exactly \eo{s}.
\end{definition}

\begin{definition}
  An inline candidate \eo{s.f a1 a2 ... aN} is \emph{inlinable}
  if the enclosing object that bounds \eo{s} has a parent
  with attribute \eo{f}.
  In this case, the parent object is called \emph{method owner} of \eo{f}.
\end{definition}

Instead of the actual inlining, in EO, we will simply replace the dynamic method call with a static reference to the said call. This is possible in EO since objects cannot be extended or modified dynamically, only decorated.

\begin{definition}
  The \emph{static form} of an inlinable candidate \eo{s.f a1 a2 ... aN}
  is produced by replacing all occurrences of \eo{s}
  with a locator referencing the method owner of \eo{f}.
\end{definition}

\begin{definition}
  A refinement of an object is its copy where zero or more inlinable candidates
  are replaced with a corresponding static form.
\end{definition}

%\begin{example}
  Consider the following object:
\begin{minted}{text,firstnumber=last,linenos,numbersep=2pt,framesep=6pt,framerule=1pt,highlightcolor=gray!30,rulecolor=gray,frame=leftline,escapeinside=||,mathescape,style=bw}
[] > a
  [self x] > f
    x > @
  [self y] > g
    [] > @
      self.f self y > a
\end{minted}
Here, the expression \eo{self.f self y} is an inline candidate.
Moreover, it is inlinable since the enclosing object that introduced \eo{self}
has a parent with method \eo{f}. Replacing the inlinable expression with its static form, we get the following:
\begin{minted}{text,firstnumber=last,linenos,numbersep=2pt,framesep=6pt,framerule=1pt,highlightcolor=gray!30,rulecolor=gray,frame=leftline,escapeinside=||,mathescape,style=bw}
[] > a
  [self x] > f
    x > @
  [self y] > g
    [] > @
      ^.^.f ^.^ y > a
\end{minted}
%\end{example}

It is important to note that a method owner can have the required attribute indirectly via decoration:

%\begin{example}
  Consider the following objects:
\begin{minted}{text,firstnumber=last,linenos,numbersep=2pt,framesep=6pt,framerule=1pt,highlightcolor=gray!30,rulecolor=gray,frame=leftline,escapeinside=||,mathescape,style=bw}
[] > a
  1 > x
  [this] > f
    this.x > @
[] > b
  a > @
  [self] > g
    self.f self > @
\end{minted}
Here, the expression \eo{self.f self} is an inlinable candidate. Note that it is inlinable since object \eo{b} decorates object \eo{a} with attribute \eo{f}.
Replacing the inlinable expression with its static form we get the following refinement of the object \eo{b}:
\begin{minted}{text,firstnumber=last,linenos,numbersep=2pt,framesep=6pt,framerule=1pt,highlightcolor=gray!30,rulecolor=gray,frame=leftline,escapeinside=||,mathescape,style=bw}
[] > b
  a > @
  [self] > g
    ^.f ^ > @
\end{minted}
%\end{example}

For the rest of this paper, we will assume that all inlineable expressions are of the form \eo{self.f self ...} where \eo{self} is always named \eo{self} and occurs in the first position as an argument to \eo{f}.

\subsubsection*{Fragile decorated objects}

Since we rely on some additional assumptions about method representation of objects, we need to specify what it means for two objects to have the same behaviour.

\begin{definition}
  Objects $x$ and $y$ are \emph{observationally equivalent} if for any expression $e$
  %\debate[Shilov]{expression <==> context in the next section?}
  that contains $x$ such that
  \begin{enumerate}
    \item $x$ does not occur in a subexpression of $e$ used as a decoratee;
    \item any method $f$ of $x$ is accessed in $e$ only in the form $x.f\; x$, making sure $x$ is passed as \eo{self} argument;
  \end{enumerate}
  evaluation of $e$ is equivalent to evaluation of an expression $e[y/x]$ (i.e. an expression $e$ where each occurrence of $x$ is replaced with $y$).
\end{definition}

Note that any object is observationally equivalent to its refinement.
We can now formulate the fragile condition for objects.

\begin{definition}
  An object $a$ is called \emph{evidently fragile} in program $P$ if there exists a refinement $a'$ of $a$ and an object $b$ that decorates, perhaps indirectly, the object $a$, such that replacing $a$ with $a'$ changes the observational properties of the object $b$.
\end{definition}

\begin{definition}
  An object $a$ is called \emph{fragile} if there exists a program $P$ such that $a$ is evidently fragile in $P$.
\end{definition}

\subsubsection*{Unanticipated mutual recursion}

For decorated objects, unanticipated recursion happens whenever it involves an inlinable candidate in a, possibly indirect, decoratee of an object. Indeed, replacing such a candidate with its static form would result in a change of control flow, which makes such mutual recursion scenario problematic.

%\begin{example}
  The following EO code has unanticipated mutual recursion:
\begin{minted}{text,firstnumber=last,linenos,numbersep=2pt,framesep=6pt,framerule=1pt,highlightcolor=gray!30,rulecolor=gray,frame=leftline,escapeinside=||,mathescape,style=bw}
[] > a
  [self] > f
    3 > @
  [self] > g
    self.f self > @

[] > b
  a > @
  [self] > f
    self.g self > @
\end{minted}
  Here, two inlineable expressions are present: \eo{self.f self} and \eo{self.g self}.
  Replacing the former with its static form changes the observational properties of object \eo{b}.
  In the original code, evaluating \eo{b.f b} would go into an infinite recursive computation.
  And after the replacement, \eo{b.f b} would result in (a decorated) object 3.
%\end{example}

% \subsubsection*{Unjustified assumptions in subclasses}
%
% To be done.

\section{Detecting unanticipated mutual recursion}
\label{section:detecting-mutual-recursion}

\subsection{Objects and contexts in Functional Notation}\label{subsection:ObjInCont}
In general, a \emph{context} is a finite set of \emph{object declarations}. This set can be expanded as EO-program (i.e., syntactically correct with a single declaration for each object). For example, below follows a context comprising two object declarations \texttt{point} and \texttt{circle} that are modified EO-objects borrowed from \cite{bugayenko2021eolang}: 
%\debate[Shilov]{Need to unify environment with {minted} throughout the section?}
\begin{minted}{text,firstnumber=last,linenos,numbersep=2pt,framesep=6pt,framerule=1pt,highlightcolor=gray!30,rulecolor=gray,frame=leftline,escapeinside=||,mathescape,style=bw}
[x y] > point
  [to] > distance
    length. > len
      vector
        to.x.sub (^.x)
        to.y.sub (^.y)
        
[center radius] > circle
  center > @
  [p] > is-inside
    (^.distance p).leq ^.radius > @
\end{minted}
  
Let us introduce the following interpretation of objects available in a given fragment.
Let us consider each object (declaration) as a \emph{parameterized recursive monadic function} (definition) where \emph{parameters} are \emph{free attributes} (of the declaration), the single argument of these functions has \texttt{string} type (that matches against \emph{bound attributes} of the declaration), and the return values of are objects (more precisely, object declarations).

Let us use the following meta-notation in this the interpretation. 
Firstly, let us use a conventional \emph{finite case-of} construct used in Mathematics, e.g. 
$\left\{
\begin{array}{l}
     \dots \mbox{, if } \dots,\\
     \dots\ \dots\ \dots \\
     \dots \mbox{ otherwise.}
\end{array} 
\right.$
Next, let us write parameters as subscripts; for example, $f_{a,b}(\mbox{\texttt{''xyz''}})_{c,d,e}$ should be understood as follows: $f_{a,b}$ is an object $f$ with two free attributes that get values $a$ and $b$; this object $f_{a,b}$ is applied (as a monadic function) to argument \texttt{''xyz''} (of the string type to be matched against bound attributes of the object) and returns an object  $f_{a,b}(\mbox{\texttt{''xyz''}})$ with three free attributes that get values $c$, $d$, and $b$.

Let us give an example. In terms of \cite{bugayenko2021eolang}, the above \emph{abstract} EO-object \texttt{circle} has two \emph{free attributes}  \texttt{center} and \texttt{radius}, a \emph{bound attribute} \texttt{is-inside}, and all attributes of the single \emph{decorating object} \texttt{center}. In our interpretation, this EO-object corresponds to the following parameterized recursive monadic function definition written in a standard mathematical notation:
\begin{equation}
\begin{array}{l}
circle_{center,\ radius}(argument)\ =\\
=\left\{
\begin{array}{l}
\left(circle_{center,\ radius}("distance")\right)_{p}("leq")_{radius}\\ \hspace*{\fill}\mbox{if}\ argument="is-inside"\\
center(argument)\ \mbox{otherwise} 
\end{array}
\right.
\end{array}\end{equation}

In general: free attributes of any EO-object become parameters' names and named bound attributes become alternative clauses in the case-of construct that ends by a default clause that corresponds to decoratee (if the object has a decoratee). Let us refer to this representation of any object declaration (i.e., any EO-object) as an  \emph{object in functional notation}; a \emph{context in functional notation} is the set of all its objects in functional notation.

\subsection{Motivating Example}\label{subsection:MotExm}
Let us consider another example that plays an important role in illustrating and explaining the suggested analysis --- the following context comprising two object declarations:
\begin{minted}{text,firstnumber=last,linenos,numbersep=2pt,framesep=6pt,framerule=1pt,highlightcolor=gray!30,rulecolor=gray,frame=leftline,escapeinside=||,mathescape,style=bw}
[] > base
  memory > x
  [self v] > n
    x.write > @
      v
  [self v] > m
    self.n > @
      self 
      v
      
[] > derived
  base > @
  [self v] > n
    self.m > @
    v
\end{minted}

This example can be rewritten in terms of parameterized recursive monadic function definition as follows:
\begin{equation}
\begin{array}{l}
base(argument)\ =\\
=\left\{
\begin{array}{l}
memory\ \mbox{if}\ argument="x"\\
x("write")_{v}\ \mbox{if}\ argument="n"\\
self("n")_{self,\ v}\ \mbox{if}\ argument="m" 
\end{array}
\right.\\
\\
derived(argument)\ =\\
=\left\{
\begin{array}{l}
self("m")_{self,\ v}\ \mbox{if}\ argument="n"\\ 
base(argument)\ \mbox{otherwise}
\end{array}
\right.
\end{array}
\end{equation}

Let us try $derived("m")_{derived}$:
\begin{equation}
\begin{gathered}
\underline{derived("n")_{derived}}\ = \hspace*{\fill}\\ 
=\ derived("m")_{derived,\ v}\ =\
 base("m")_{derived,\ v}\ =\\
\hspace*{\fill}=\ \underline{derived("n")_{derived}}\ _{,v}
\end{gathered}
\end{equation}
i.e. we have infinite recursion (which most probably is  \emph{unintended/unanticipated}).

\subsection{Analysis Method for  Contexts}\label{subsection:Method}
An idea behind the method below is a loop search in  the call graph of a context abstracted as a \underline{deterministic} \underline{finite automaton}, where states are EO-objects with the first free attributes (i.e., parameterized functions with the first parameters in functional notation) and transition rules are calls in the call graph (i.e., all clauses of the context).  

\subsubsection{Method in the functional notation} 
Let us start with method description assuming that we are given a context  (a program in particular) in the functional notation.
\begin{enumerate}
    \item Let $n$ be the total number of clauses in the context.
    \item For each object $obj$ in the context, for each clause in its declaration that has a pattern $$prm(val^{\prime})_{prm\dots}\ \mbox{if}\ arg=val$$ exercise symbolically a call  $obj(val)_{obj}$ until the next instance of the $obj(val)_{obj\dots}$ (if any), but not more than $n$ calls in the row:
    \begin{itemize}
        \item if the exercise has the next instance, then make a warning about a possible infinite recursion and report the exercise.
    \end{itemize}
    \item For each object $obj$ in the context, for each default clause in its declaration that has a pattern $$prm(val^{\prime})_{prm\dots}\ \mbox{otherwise},$$ for each attribute $val$ that is bound anywhere in the context but not in the declaration, exercise symbolically a call  $obj(val)_{obj}$ until the next instance of the $obj(val)_{obj\dots}$ (if any), but not more than $n$ calls in the row:
    \begin{itemize}
        \item if the exercise has the next instance, then make a warning about a possible infinite recursion and report the exercise.
    \end{itemize}
\end{enumerate}

\subsubsection{Method in EO-terms} 
Now let us present the method description assuming that we are given any EO-context.
\begin{enumerate}
    \item Let $n$ be the total number of clauses in the context.
    \item For each object $obj$ in the context, for each clause in its declaration that has a pattern 
    \begin{equation}
    \begin{array}{l}
    [\dots]\ >\ obj \\
    \hspace*{2em} \dots \\
    \hspace*{2em} [prm\ \dots]\ >\ atr \\
    \hspace*{2em} \dots
    \end{array}
    \end{equation}
    exercise symbolically a call  $obj.atr\ obj$ until the next instance of the $obj.atr\ obj$ (if any), but not more than $n$ calls in the row:
    \begin{itemize}
        \item if the exercise has the next instance, then make a warning about a possible infinite recursion and report the exercise.
    \end{itemize}
    \item For each object $obj$ in the context, for each default clause in its declaration that has a pattern
    \begin{equation}
    \begin{array}{l}
    [...]\ >\ obj \\
    \hspace*{2em} \dots \\
    \hspace*{2em} [prm\ \dots]\ >\ @ \\
    \hspace*{2em} \dots
    \end{array}
    \end{equation}
    for each attribute $atr$ that is bound anywhere in the context but not in the object, exercise symbolically a call  $obj.atr\ obj$ until the next instance of the $obj.atr\ obj$ (if any), but not more than $n$ calls in the row:
    \begin{itemize}
        \item if the exercise has the next instance, then make a warning about a possible infinite recursion and report the exercise.
    \end{itemize}
\end{enumerate}

\subsubsection{Comments:}  
\begin{enumerate}
  \item The time complexity of the analysis is $O(n^2)$ where $n$ is the total number of clauses in a given context (because the analysis is just a loop analysis of a finite deterministic automaton with $n$ states).
  \item The above analysis over-approximate the set of possible infinite recursive loops in particular because some of these loops may be non-reachable from the main method of a given program; better (more accurate) analysis should be based on \emph{lasso detection} in finite automata (i.e.,  finite legal executions where some state repeats twice)  \cite{mudduluru2017lasso}. 
\end{enumerate}

\subsection{Examples}
Firstly, we would like to refer to the motivating example given in subsection \ref{subsection:MotExm}. A more complicated example is presented below:
\begin{minted}{text,firstnumber=last,linenos,numbersep=2pt,framesep=6pt,framerule=1pt,highlightcolor=gray!30,rulecolor=gray,frame=leftline,escapeinside=||,mathescape,style=bw}
[] > base
  memory > x
  [self v] > n
    x.write > @
      v
  [self v] > m
    self.n > @
      self 
      v
      
[] > derived
  base > @
  [self v] > o
    self.m > @
      self
      v
      
[] > derived_again
  derived > @
  [self v] > n
    self.o > @
      self
      v
\end{minted}

A couple of examples of infinite recursion (detected by the method described in the subsection \ref{subsection:Method}) follows:
\begin{itemize}
    \item \texttt{derived.o} $\rightarrow$ \texttt{base.m} $\rightarrow$ \texttt{derived\_again.n} $\rightarrow$ \texttt{derived.o}
    \item \texttt{derived\_again.n} $\rightarrow$ \texttt{derived.o} $\rightarrow$ \texttt{base.m} $\rightarrow$ \texttt{derived\_again.n}
\end{itemize}

\section{Implementation}
\label{section:implementation}
The algorithm described in section 4 was implemented in the programming language Scala with the \textit{cats} library \cite{scalacats}. The implementation can be described as a series of transformations on EO source code, also known as \textit{passes}. In this chapter, we will show how the analysis is performed on the EO code in figure \ref{fig:implementation-example}:

\begin{figure}[ht]
    \centering
\begin{minted}{text,firstnumber=last,linenos,numbersep=2pt,framesep=6pt,framerule=1pt,highlightcolor=gray!30,rulecolor=gray,frame=leftline,escapeinside=||,mathescape,style=bw}
[] > a
  [] > new
    b.new > @
    [self x y] > f
      self.g self y x > @

[] > b
  [] > new
    c.new > @
[] > c
  [] > new
    [self y x] > g
      self.f self x y > @
    [self x y] > f
      self > @
\end{minted}
    \caption{Example EO program.}
    \label{fig:implementation-example}
\end{figure}

\subsection{Terminology}

A \textit{method} is an abstract EO object with at least one free attribute. The first free attribute of this object has the name \textit{self}. When the method is called, it is assumed that the calling object is passed to the self attribute. In figure \ref{fig:implementation-example}, EO objects "a.new.f", "c.new.f", and "c.new.g" are methods.

An \textit{object} is a shorthand for an abstract EO object without free attributes. "a", "b", "c", "a.new", "b.new", "c.new" in figure \ref{fig:implementation-example} can be called just objects.

\subsection{Preprocessing} 
EO source code is first parsed into an abstract syntax tree (AST), which is then transformed into a \textit{partially-resolved object tree}. It is a tree-like intermediate data structure that reflects the nesting structure of the program. % add more
Each node of this tree represents an EO object and contains the following information:

\begin{itemize}
    \item Fully-qualified name of the object. % derived
    \item The name of the decorated object. % base
    \item An association between the method names defined in the object and the methods they call. 
\end{itemize}

The last two items together can be described as a \textit{partial call graph} of the object. It is called \textit{partial} because it holds only the names of the methods, but not the objects they come from. The methods may be defined in the same object where they are called, as well as in the decorated object. The decorated object, in turn, can "inherit" methods from its own decorated object, etc. Resolving the closest decorated object where the method was last redefined is the objective of the next pass.

\begin{figure}[ht]
    \centering
    \normalsize
    \includegraphics[width=0.45\textwidth, height=0.25\textwidth]{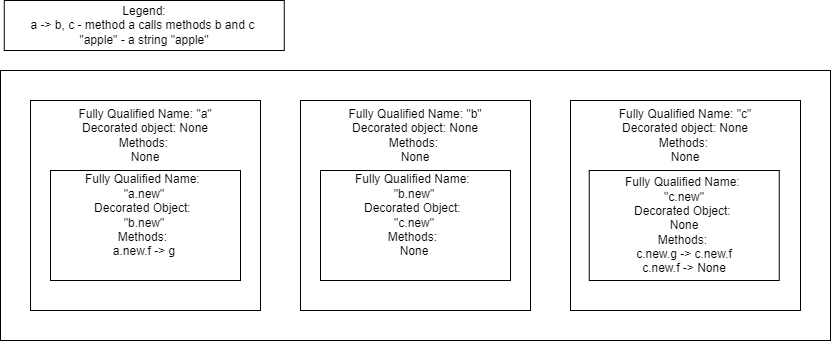}
    \caption{Object tree after the preprocessing step.}
    \label{fig:preprocessing}
\end{figure}

After this step, the EO code \ref{fig:implementation-example} would be transformed into the tree in figure \ref{fig:preprocessing}.

\subsection{Resolving Decorated Objects}
The next pass transforms the \textit{partially-resolved object tree} into a \textit{resolved object tree}, where all the information about the decorated objects is filled in. The overall structure of this tree remains the same; however the node structure changes:

\begin{itemize}
    \item All the \textit{partially-resolved} calls now contain the information about the last object where they were redefined.
    \item The name of the decorated object is replaced with a reference to the decorated object itself.
    \item The call graph of the object is extended with the methods that come from the decorated object.
\end{itemize}

\begin{figure}[ht]
    \centering
    \includegraphics[width=0.45\textwidth, height=0.25\textwidth]{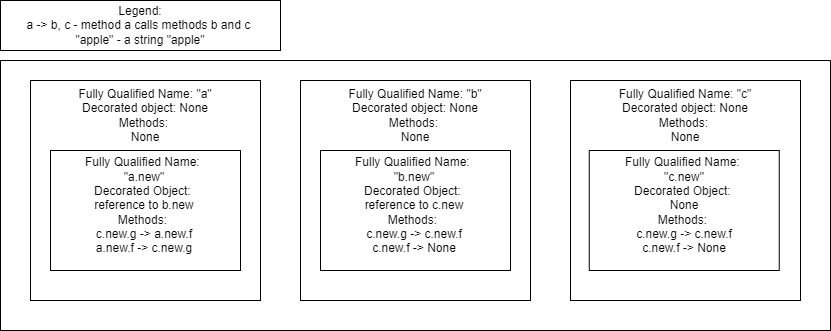}
    \caption{Object tree with the resolved decorated objects.}
    \label{fig:resolving-decorated-objects}
\end{figure}

Figure \ref{fig:resolving-decorated-objects} shows what the object tree of program \ref{fig:implementation-example} will look like after the first 2 passes. It is worth mentioning that object "b.new", which previously had no method defined, has all the methods that come from its decorated object "c.new". As for object "a.new", method "g" that is called by method "f" is correctly resolved to come from the decorated object "c.new". Even though it technically comes from the object "b.new", "c.new" is the last object where it was \textit{redefined}.

\subsection{Analysis}
The final pass is essentially a traversal of the \textit{resolved object tree}. For each object node of the tree, the extended call graph is traversed in a depth-first manner to find the call chains that span multiple objects. When the traversal encounters the method that is already present in the call chain, it is considered a call cycle and the traversal stops. 

The results of such traversals for each object are accumulated and presented to the user as console messages. For the example program in figure \ref{fig:implementation-example}, the message produced will be the following:
\newline
\begin{mdframed}
    \begin{minted}{text}
a.new: 
  a.new.g (was last redefined in "c.new") -> 
  a.new.f -> 
  a.new.g (was last redefined in "c.new")
    \end{minted}
\end{mdframed}

This means that for the object "a.new", there exists a never-ending call-chain: method "g" (which is last redefined in "c.new") calls method "f", which, in turn calls method "g" again, etc.

\section{Benchmarking methodology}
\label{section:benchmarking-methodology}

In this section, we describe our approach to benchmarking static analysis tools for C++ and EO. We first describe the general methodology, then the format of the input-output data. Finally, we give a description of the set of metrics based on a part of the resulting report.

\subsection{Comparing static analysis tools}

In this paper, we use a direct and simple approach to comparing static analysis tools. Put simply, we have a collection of example programs marked as \emph{good} (meaning that the program is defect-free) or \emph{bad} (meaning that it has some defect). We run static analysis tools on these programs and check whether the tool agrees with the markings.

The approach has several limitations, such as ignoring the actual type, location, and confidence level of defects reported by the tool, as well as supporting programs with multiple defects. However, for preliminary comparison, this approach works well.

To organize the comparison, we collected a suite of test files for each type of defect.
Each test file targets a specific circumstance of the defect. For instance, mutual recursion may be harder to detect for some tools in a complicated hierarchy of inheritance or long chain of calls, so we should add test files for such scenarios.

Each test file, for each supported programming language, presents two similar versions of a program~--- one \emph{good} and one \emph{bad}. The versions of the programs can be thought of as ``before'' and ``after'' fixing the corresponding defect. The versions in different languages are expected to be equivalent, at least from a software engineer's perspective. The translation from C++ to EO was done manually since, at the time of writing the article, the \texttt{c2eo} translator was still in development and could not cover all features from our test suit.

\begin{figure}[ht]
    \centering
    \begin{mdframed}
\begin{minted}{yaml}
title: # Title
description: >
  # Detailed description
features: # a list of tags
bad:
  source.cpp: |
    # bad C++ program
  test.eo: |
    # bad EO program
good:
  source.cpp: |
    # good C++ program
  test.eo: |
    # good EO program
\end{minted}
    \end{mdframed}
    \caption{Test file structure in YAML format.}
    \label{fig:test-file}
\end{figure}

We implement test files as YAML documents with structure as shown in Figure~\ref{fig:test-file}. Such YAML files are then used by automatic continuous integration scripts to evaluate static analyzers whenever the benchmark suite repository on GitHub is updated.

\begin{table*}[ht]
    \caption{Comparison of performance metrics for Polystat and Clang-Tidy for unanticipated mutual recursion defect.}
    \label{table:metrics}
    \centering
    \normalsize
    \begin{tabular}{|l|l|r|r|r|r|r|r|r|r|r|}%
\hline%
Analyzer&Defect title&TP&TN&FP&FN&ERR&Accuracy&Precision&Recall&F1 score\\%
\hline%
\hline%
Clang{-}Tidy&Mutual recursion&0&26&0&26&0&50.0\%&0.0\%&0.0\%&0.0\%\\%
\hline%
Polystat&Mutual recursion&26&22&4&0&0&92.3\%&86.7\%&100.0\%&92.9\%\\%
\hline%
    \end{tabular}%
\end{table*}

Assuming each tool is evaluated using one programming language, every test file contains essentially two programs. Running a tool on a program leads to one of the following possible outcomes:
\begin{enumerate}
    \item True Positive(TP)~--- defect is detected in a \emph{bad} program;
    \item False Positive(FP)~--- defect is detected in a \emph{good} program;
    \item True Negative(TN)~--- no defect detected in a \emph{good} program;
    \item False Negative(FN)~--- no defect detected in a \emph{bad} program;
    \item Error(ERR)~--- the tool exited with a non-zero exit code or crashed.
\end{enumerate}

Evaluating each tool on a collection of test files, we accumulate the following metrics:
\begin{itemize}
    \item Total count per type of outcome (TP, FP, TN, FN, ERR);
    \item \emph{Accuracy}~-- a ratio of \emph{true} outcomes to the total number of test programs; this metric helps understand how good a tool is at predicting the presence of a defect;
    \item \emph{Precision}~--- a ratio of \emph{true positives} to the total number of \emph{positive} outcomes (predicted positives); this metric helps us understand how ``useful'' are positive detections of a defect in a program by a tool;
    \item \emph{Recall}~--- a ratio of \emph{true positives} to the total number of \emph{bad} programs (actual positives); this metric helps us understand how well actual defects are detected by a tool;
    \item \emph{F1 score}~--- a harmonic mean of Precision and Recall; this metric is commonly used for preliminary comparison of tools, a high F1 score indicates that both Precision and Recall are good.
\end{itemize}

\begin{figure}[ht]
    \centering
    \begin{mdframed}
    \begin{align}
        \mathsf{Accuracy} &=\frac{TP + TN} {TP + TN + FP + FN + ERR} \\
        \mathsf{Precision} &= \frac{TP} {TP + FP} \\
        \mathsf{Recall} &= \frac{TP} {TP + FN} \\
        \mathsf{F1} &= \frac{2 \times \mathsf{Precision} \times \mathsf{Recall}} {\mathsf{Precision} + \mathsf{Recall}}
    \end{align}
    \end{mdframed}
    \caption{Basic performance metrics for static analysis tools.}
    \label{fig:metrics}
\end{figure}

Executing the overall benchmark produces an automated report consisting of three parts. All metrics are grouped by the type of defect and the tool forming the \emph{statistics table}. A detailed account of specific output for each tool run of every test is recorded in a \emph{details table}. All defect detection outputs are grouped by tool and presented in \emph{detection messages}. In this paper, we present only the statistics table, leaving the intricate parts of the report in the appendix.

\section{Benchmarking results}
\label{section:benchmarking-results}

Benchmarking was carried out on a set of \emph{52} tests. In Table~\ref{table:metrics}, we can see a comparison of Polystat and Clang-Tidy performance for unanticipated mutual recursion defect.

Polystat has an accuracy of 92\% for detecting unanticipated mutual recursion. Note that the recall here is 100\%, meaning that Polystat has successfully detected all \emph{bad} programs in test files. However, because of some false negatives, the overall accuracy is not as high. The false negatives come from programs with branching (such as \ff{if} statements or \ff{while} loops), and Polystat in its current state cannot properly understand whether the condition should be taken into account. Still, Polystat demonstrates a 42\% improvement over Clang-Tidy for this type of defect.

Clang-Tidy does not claim and indeed does not find any of the OOP-specific defects. However, since it does not have false positives for those cases, the accuracy is exactly 50\%.
\section{Discussion and future work}
\label{section:discussion}

We have presented a plausible approach to translating object-oriented programs in Java, C++, and Python to a common intermediate representation~--- the EO programming language. In this paper, we have limited ourselves to a minimal translation that captures the main structural properties of classes, inheritance, and method overriding. This approach works well for the purpose of analysis in this paper but is incomplete for richer analysis and other applications, such as compiling or interpreting via EO intermediate representation. More research has to be performed towards a more faithful translation.

This paper defines refinements of classes and objects in terms of inlining or fixing method calls statically. Such a definition allows the automatic generation of many refinements, testing them, or even verifying some of their observational properties. We believe that this limited notion should work well in practice for fragile base class analysis. Mikhajlov and Sekerinski \cite{MikhajlovSekerinski1998} provide a more general description of refinement, which is worth exploring and adapting to Elegant Objects in further research.

Mikhajlov and Sekerinski's work \cite{MikhajlovSekerinski1998} also considers the original version of a base class and its refinement as inputs for analysis. In contrast, we choose to operate with a single version assuming that any refinements may be applied in the development process. A possible extension of our research may concern the analysis of evolving EO code for a more precise analysis.

We can formulate a specialized fragile base class problem with a specific observational property of interest. Introducing types or contracts in EO in the form of pre- and post-conditions for methods, invariants of objects, or other properties should help generalize and automate fragile base class analysis.

Types, in particular, can be used to decrease false positives for unanticipated mutual recursion detection, ignoring method calls that should not be possible because of type error. Types that support automatic type inference, such as row types with row polymorphism \cite{Wand1991}, should be a good candidate for further research in this direction.

Class invariants, as well as pre- and post-conditions of methods, could also be automatically inferred, at least partially, relying on techniques, such as Logozzo's \cite{Logozzo2004}.

\section{Conclusion}

We have presented a method for detecting a certain subset of fragile base class problems in object-oriented programs via EO language. We have shown that a translation to EO is possible for the mainstream object-oriented languages, such as Java, C++, and Python. Moreover, such a translation is faithful with respect to the fragile base class problem~--- translated programs have fragile decorated objects if and only if the original program has fragile base classes.

We have noted the limitations of EO as an intermediate language for static analysis. In particular, we suggested extending EO with standardized metadata and types to preserve more information from the source program. Such extensions, in turn, would allow a more accurate analysis, especially when pre- and post-conditions are involved.

% \begin{acks}
\subsubsection{Acknowledgements}
  This research has been generously funded by Huawei in the framework of Polystat project.
  We thank Yegor Bugayenko for taking his time to explain the ideas behind EO.
  We also thank Tymur Lysenko who has worked on an early prototype of a static analyzer based loosely on ideas we describe in this paper. Finally, we thank Violetta Sim, Mansur Khazeev, and Ruslan Saduov for proofreading the paper.
% \end{acks}

% \bibliographystyle{ACM-Reference-Format}
\bibliographystyle{splncs04}
\bibliography{sample-base}

% \onecolumn

\appendix
\section*{Statistic table}%
\label{sec:Statistictable}%
\renewcommand{\arraystretch}{1.25}%
\begin{tabular}{|l|l|r|r|r|r|r|r|r|r|r|}%
\hline%
Analyzer&Defect title&TP&TN&FP&FN&ERR&Accuracy&Precision&Recall&F1 score\\%
\hline%
\hline%
Polystat (EO)&mutual{-}recursion&26&22&4&0&0&92.3\%&86.7\%&100.0\%&92.9\%\\%
\hline%
\textbf{Polystat (EO)}&\textbf{All}&\textbf{26}&\textbf{22}&\textbf{4}&\textbf{0}&\textbf{0}&\textbf{92.3\%}&\textbf{86.7\%}&\textbf{100.0\%}&\textbf{92.9\%}\\%
\hline%
\hline%
Clang{-}Tidy&mutual{-}recursion&0&26&0&26&0&50.0\%&0.0\%&0.0\%&0.0\%\\%
\hline%
\textbf{Clang{-}Tidy}&\textbf{All}&\textbf{0}&\textbf{26}&\textbf{0}&\textbf{26}&\textbf{0}&\textbf{50.0\%}&\textbf{0.0\%}&\textbf{0.0\%}&\textbf{0.0\%}\\%
\hline%
\end{tabular}%

\medskip Description%
\begin{itemize}%
\item%
True Positive(TP) {-} warnings exist and should be%
\item%
True Negative(TN) {-} no warnings and shouldn't be%
\item%
False Negative(FN) {-} no warnings, but they should be%
\item%
False Positive(FP) {-} warnings exist but shouldn't be%
\item%
Errors(ERR) {-} errors/exceptions during analysis%
\end{itemize}

\section*{Details table}%
\label{sec:Detailstable}%
\renewcommand{\arraystretch}{1.25}%
%\begin{landscape}
\begin{longtable}{|l|c|c|}%
\hline%
\multicolumn{1}{|c|}{File}&Polystat (EO)&Clang{-}Tidy\\%
\hline%
\href{https://github.com/Polystat/awesome-bugs/blob/master/tests/inheritance/mutual-recursion-in-chain-of-calls.yml}{inheritance/mutual{-}recursion{-}in{-}chain{-}of{-}calls.yml}&OK&FN\\%
\hline%
\href{https://github.com/Polystat/awesome-bugs/blob/master/tests/inheritance/mutual-recursion-in-factory.yml}{inheritance/mutual{-}recursion{-}in{-}factory.yml}&OK&FN\\%
\hline%
\href{https://github.com/Polystat/awesome-bugs/blob/master/tests/inheritance/mutual-recursion-in-inheritance-chain-1.yml}{inheritance/mutual{-}recursion{-}in{-}inheritance{-}chain{-}1.yml}&OK&FN\\%
\hline%
\href{https://github.com/Polystat/awesome-bugs/blob/master/tests/inheritance/mutual-recursion-in-inheritance-chain-2.yml}{inheritance/mutual{-}recursion{-}in{-}inheritance{-}chain{-}2.yml}&OK&FN\\%
\hline%
\href{https://github.com/Polystat/awesome-bugs/blob/master/tests/inheritance/mutual-recursion-in-inheritance-chain-3.yml}{inheritance/mutual{-}recursion{-}in{-}inheritance{-}chain{-}3.yml}&OK&FN\\%
\hline%
\href{https://github.com/Polystat/awesome-bugs/blob/master/tests/inheritance/mutual-recursion-in-inheritance-chain-4.yml}{inheritance/mutual{-}recursion{-}in{-}inheritance{-}chain{-}4.yml}&OK&FN\\%
\hline%
\href{https://github.com/Polystat/awesome-bugs/blob/master/tests/inheritance/mutual-recursion-in-inheritance-chain-nested-1.yml}{inheritance/mutual{-}recursion{-}in{-}inheritance{-}chain{-}nested{-}1.yml}&OK&FN\\%
\hline%
\href{https://github.com/Polystat/awesome-bugs/blob/master/tests/inheritance/mutual-recursion-in-inheritance-chain-nested-2.yml}{inheritance/mutual{-}recursion{-}in{-}inheritance{-}chain{-}nested{-}2.yml}&OK&FN\\%
\hline%
\href{https://github.com/Polystat/awesome-bugs/blob/master/tests/inheritance/mutual-recursion-in-inheritance-chain-nested-3.yml}{inheritance/mutual{-}recursion{-}in{-}inheritance{-}chain{-}nested{-}3.yml}&OK&FN\\%
\hline%
\href{https://github.com/Polystat/awesome-bugs/blob/master/tests/inheritance/mutual-recursion-in-inheritance-chain-nested-4.yml}{inheritance/mutual{-}recursion{-}in{-}inheritance{-}chain{-}nested{-}4.yml}&OK&FN\\%
\hline%
\href{https://github.com/Polystat/awesome-bugs/blob/master/tests/inheritance/mutual-recursion-in-inheritance-chain-nested-base-1.yml}{inheritance/mutual{-}recursion{-}in{-}inheritance{-}chain{-}nested{-}base{-}1.yml}&OK&FN\\%
\hline%
\href{https://github.com/Polystat/awesome-bugs/blob/master/tests/inheritance/mutual-recursion-in-inheritance-chain-nested-base-2.yml}{inheritance/mutual{-}recursion{-}in{-}inheritance{-}chain{-}nested{-}base{-}2.yml}&OK&FN\\%
\hline%
\href{https://github.com/Polystat/awesome-bugs/blob/master/tests/inheritance/mutual-recursion-in-inheritance-chain-nested-base-3.yml}{inheritance/mutual{-}recursion{-}in{-}inheritance{-}chain{-}nested{-}base{-}3.yml}&OK&FN\\%
\hline%
\href{https://github.com/Polystat/awesome-bugs/blob/master/tests/inheritance/mutual-recursion-in-inheritance-chain-nested-base-4.yml}{inheritance/mutual{-}recursion{-}in{-}inheritance{-}chain{-}nested{-}base{-}4.yml}&OK&FN\\%
\hline%
\href{https://github.com/Polystat/awesome-bugs/blob/master/tests/inheritance/mutual-recursion-in-inheritance-chain-nested-derived-1.yml}{inheritance/mutual{-}recursion{-}in{-}inheritance{-}chain{-}nested{-}derived{-}1.yml}&OK&FN\\%
\hline%
\href{https://github.com/Polystat/awesome-bugs/blob/master/tests/inheritance/mutual-recursion-in-inheritance-chain-nested-derived-2.yml}{inheritance/mutual{-}recursion{-}in{-}inheritance{-}chain{-}nested{-}derived{-}2.yml}&OK&FN\\%
\hline%
\href{https://github.com/Polystat/awesome-bugs/blob/master/tests/inheritance/mutual-recursion-in-inheritance-chain-nested-derived-3.yml}{inheritance/mutual{-}recursion{-}in{-}inheritance{-}chain{-}nested{-}derived{-}3.yml}&OK&FN\\%
\hline%
\href{https://github.com/Polystat/awesome-bugs/blob/master/tests/inheritance/mutual-recursion-in-inheritance-chain-nested-derived-4.yml}{inheritance/mutual{-}recursion{-}in{-}inheritance{-}chain{-}nested{-}derived{-}4.yml}&OK&FN\\%
\hline%
\href{https://github.com/Polystat/awesome-bugs/blob/master/tests/inheritance/mutual-recursion-nested-base.yml}{inheritance/mutual{-}recursion{-}nested{-}base.yml}&OK&FN\\%
\hline%
\href{https://github.com/Polystat/awesome-bugs/blob/master/tests/inheritance/mutual-recursion-nested-derived.yml}{inheritance/mutual{-}recursion{-}nested{-}derived.yml}&OK&FN\\%
\hline%
\href{https://github.com/Polystat/awesome-bugs/blob/master/tests/inheritance/mutual-recursion-nested.yml}{inheritance/mutual{-}recursion{-}nested.yml}&OK&FN\\%
\hline%
\href{https://github.com/Polystat/awesome-bugs/blob/master/tests/inheritance/mutual-recursion-with-if-branching1.yml}{inheritance/mutual{-}recursion{-}with{-}if{-}branching1.yml}&FP&FN\\%
\hline%
\href{https://github.com/Polystat/awesome-bugs/blob/master/tests/inheritance/mutual-recursion-with-if-branching2.yml}{inheritance/mutual{-}recursion{-}with{-}if{-}branching2.yml}&FP&FN\\%
\hline%
\href{https://github.com/Polystat/awesome-bugs/blob/master/tests/inheritance/mutual-recursion-with-if-branching3.yml}{inheritance/mutual{-}recursion{-}with{-}if{-}branching3.yml}&FP&FN\\%
\hline%
\href{https://github.com/Polystat/awesome-bugs/blob/master/tests/inheritance/mutual-recursion-with-random-if-branching.yml}{inheritance/mutual{-}recursion{-}with{-}random{-}if{-}branching.yml}&FP&FN\\%
\hline%
\href{https://github.com/Polystat/awesome-bugs/blob/master/tests/inheritance/mutual-recursion.yml}{inheritance/mutual{-}recursion.yml}&OK&FN\\%
\hline%
\end{longtable}%
%\end{landscape}

\medskip Description%
\begin{itemize}%
\item%
OK = TP and PN%
\item%
FN = FN and TP%
\item%
FP = FP and TN%
\item%
FF = FP and FN%
\item%
E {-} errors/exceptions during analysis%
\end{itemize}

\section*{Detected defect details}%
\label{sec:Detecteddefectdetails}%
\subsection*{Polystat}%
\label{subsec:Polystat}%
\begin{enumerate}%
\item%
temp/sources/eo/mutual{-}recursion{-}bad: test.derived: test.derived.m (was last redefined in "test.base") {-}> test.derived.n {-}> test.derived.m (was last redefined in "test.base")%
\item%
temp/sources/eo/mutual{-}recursion{-}in{-}chain{-}of{-}calls{-}bad: test.derived: test.derived.o (was last redefined in "test.base") {-}> test.derived.n {-}> test.derived.m (was last redefined in "test.base") {-}> test.derived.o (was last redefined in "test.base")%
\item%
temp/sources/eo/mutual{-}recursion{-}in{-}factory{-}bad: test.derived: test.derived.m (was last redefined in "test.base\_factory.get\_base") {-}> test.derived.n {-}> test.derived.m (was last redefined in "test.base\_factory.get\_base")%
\item%
temp/sources/eo/mutual{-}recursion{-}in{-}inheritance{-}chain{-}1{-}bad: test.derived: test.derived.m (was last redefined in "test.base") {-}> test.derived.n {-}> test.derived.m (was last redefined in "test.base")\newline%
test.derived\_again: test.derived\_again.m (was last redefined in "test.base") {-}> test.derived\_again.n (was last redefined in "test.derived") {-}> test.derived\_again.m (was last redefined in "test.base")%
\item%
temp/sources/eo/mutual{-}recursion{-}in{-}inheritance{-}chain{-}2{-}bad: test.derived\_again: test.derived\_again.m (was last redefined in "test.base") {-}> test.derived\_again.n {-}> test.derived\_again.m (was last redefined in "test.base")%
\item%
temp/sources/eo/mutual{-}recursion{-}in{-}inheritance{-}chain{-}3{-}bad: test.derived\_again: test.derived\_again.m (was last redefined in "test.derived") {-}> test.derived\_again.n {-}> test.derived\_again.m (was last redefined in "test.derived")%
\item%
temp/sources/eo/mutual{-}recursion{-}in{-}inheritance{-}chain{-}4{-}bad: test.derived\_again: test.derived\_again.m (was last redefined in "test.base") {-}> test.derived\_again.n {-}> test.derived\_again.o (was last redefined in "test.derived") {-}> test.derived\_again.m (was last redefined in "test.base")%
\item%
temp/sources/eo/mutual{-}recursion{-}in{-}inheritance{-}chain{-}nested{-}1{-}bad: test.very\_outer.outer.derived: test.very\_outer.outer.derived.m (was last redefined in "test.very\_outer.outer.base") {-}> test.very\_outer.outer.derived.n {-}> test.very\_outer.outer.derived.m (was last redefined in "test.very\_outer.outer.base")\newline%
test.very\_outer.outer.derived\_again: test.very\_outer.outer.derived\_again.m (was last redefined in "test.very\_outer.outer.base") {-}> test.very\_outer.outer.derived\_again.n (was last redefined in "test.very\_outer.outer.derived") {-}> test.very\_outer.outer.derived\_again.m (was last redefined in "test.very\_outer.outer.base")%
\item%
temp/sources/eo/mutual{-}recursion{-}in{-}inheritance{-}chain{-}nested{-}2{-}bad: test.very\_outer.outer.derived\_again: test.very\_outer.outer.derived\_again.m (was last redefined in "test.very\_outer.outer.base") {-}> test.very\_outer.outer.derived\_again.n {-}> test.very\_outer.outer.derived\_again.m (was last redefined in "test.very\_outer.outer.base")%
\item%
temp/sources/eo/mutual{-}recursion{-}in{-}inheritance{-}chain{-}nested{-}3{-}bad: test.very\_outer.outer.derived\_again: test.very\_outer.outer.derived\_again.m (was last redefined in "test.very\_outer.outer.derived") {-}> test.very\_outer.outer.derived\_again.n {-}> test.very\_outer.outer.derived\_again.m (was last redefined in "test.very\_outer.outer.derived")%
\item%
temp/sources/eo/mutual{-}recursion{-}in{-}inheritance{-}chain{-}nested{-}4{-}bad: test.very\_outer.outer.derived\_again: test.very\_outer.outer.derived\_again.m (was last redefined in "test.very\_outer.outer.base") {-}> test.very\_outer.outer.derived\_again.n {-}> test.very\_outer.outer.derived\_again.o (was last redefined in "test.very\_outer.outer.derived") {-}> test.very\_outer.outer.derived\_again.m (was last redefined in "test.very\_outer.outer.base")%
\item%
temp/sources/eo/mutual{-}recursion{-}in{-}inheritance{-}chain{-}nested{-}base{-}1{-}bad: test.derived: test.derived.m (was last redefined in "test.very\_outer.outer.base") {-}> test.derived.n {-}> test.derived.m (was last redefined in "test.very\_outer.outer.base")\newline%
test.derived\_again: test.derived\_again.m (was last redefined in "test.very\_outer.outer.base") {-}> test.derived\_again.n (was last redefined in "test.derived") {-}> test.derived\_again.m (was last redefined in "test.very\_outer.outer.base")%
\item%
temp/sources/eo/mutual{-}recursion{-}in{-}inheritance{-}chain{-}nested{-}base{-}2{-}bad: test.derived\_again: test.derived\_again.m (was last redefined in "test.very\_outer.outer.base") {-}> test.derived\_again.n {-}> test.derived\_again.m (was last redefined in "test.very\_outer.outer.base")%
\item%
temp/sources/eo/mutual{-}recursion{-}in{-}inheritance{-}chain{-}nested{-}base{-}3{-}bad: test.derived\_again: test.derived\_again.m (was last redefined in "test.derived") {-}> test.derived\_again.n {-}> test.derived\_again.m (was last redefined in "test.derived")%
\item%
temp/sources/eo/mutual{-}recursion{-}in{-}inheritance{-}chain{-}nested{-}base{-}4{-}bad: test.derived\_again: test.derived\_again.m (was last redefined in "test.very\_outer.outer.base") {-}> test.derived\_again.n {-}> test.derived\_again.o (was last redefined in "test.derived") {-}> test.derived\_again.m (was last redefined in "test.very\_outer.outer.base")%
\item%
temp/sources/eo/mutual{-}recursion{-}in{-}inheritance{-}chain{-}nested{-}derived{-}1{-}bad: test.very\_outer.outer.derived: test.very\_outer.outer.derived.m (was last redefined in "test.base") {-}> test.very\_outer.outer.derived.n {-}> test.very\_outer.outer.derived.m (was last redefined in "test.base")\newline%
test.very\_outer.outer.derived\_again: test.very\_outer.outer.derived\_again.m (was last redefined in "test.base") {-}> test.very\_outer.outer.derived\_again.n (was last redefined in "test.very\_outer.outer.derived") {-}> test.very\_outer.outer.derived\_again.m (was last redefined in "test.base")%
\item%
temp/sources/eo/mutual{-}recursion{-}in{-}inheritance{-}chain{-}nested{-}derived{-}2{-}bad: test.very\_outer.outer.derived\_again: test.very\_outer.outer.derived\_again.m (was last redefined in "test.base") {-}> test.very\_outer.outer.derived\_again.n {-}> test.very\_outer.outer.derived\_again.m (was last redefined in "test.base")%
\item%
temp/sources/eo/mutual{-}recursion{-}in{-}inheritance{-}chain{-}nested{-}derived{-}3{-}bad: test.very\_outer.outer.derived\_again: test.very\_outer.outer.derived\_again.m (was last redefined in "test.very\_outer.outer.derived") {-}> test.very\_outer.outer.derived\_again.n {-}> test.very\_outer.outer.derived\_again.m (was last redefined in "test.very\_outer.outer.derived")%
\item%
temp/sources/eo/mutual{-}recursion{-}in{-}inheritance{-}chain{-}nested{-}derived{-}4{-}bad: test.very\_outer.outer.derived\_again: test.very\_outer.outer.derived\_again.m (was last redefined in "test.base") {-}> test.very\_outer.outer.derived\_again.n {-}> test.very\_outer.outer.derived\_again.o (was last redefined in "test.very\_outer.outer.derived") {-}> test.very\_outer.outer.derived\_again.m (was last redefined in "test.base")%
\item%
temp/sources/eo/mutual{-}recursion{-}nested{-}bad: test.very\_outer.outer.derived: test.very\_outer.outer.derived.m (was last redefined in "test.very\_outer.outer.base") {-}> test.very\_outer.outer.derived.n {-}> test.very\_outer.outer.derived.m (was last redefined in "test.very\_outer.outer.base")%
\item%
temp/sources/eo/mutual{-}recursion{-}nested{-}base{-}bad: test.derived: test.derived.m (was last redefined in "test.very\_outer.outer.base") {-}> test.derived.n {-}> test.derived.m (was last redefined in "test.very\_outer.outer.base")%
\item%
temp/sources/eo/mutual{-}recursion{-}nested{-}derived{-}bad: test.very\_outer.outer.derived: test.very\_outer.outer.derived.m (was last redefined in "test.base") {-}> test.very\_outer.outer.derived.n {-}> test.very\_outer.outer.derived.m (was last redefined in "test.base")%
\item%
temp/sources/eo/mutual{-}recursion{-}with{-}if{-}branching1{-}bad: test.derived: test.derived.m (was last redefined in "test.base") {-}> test.derived.n {-}> test.derived.m (was last redefined in "test.base")%
\item%
temp/sources/eo/mutual{-}recursion{-}with{-}if{-}branching1{-}good: test.derived: test.derived.m (was last redefined in "test.base") {-}> test.derived.n {-}> test.derived.m (was last redefined in "test.base")%
\item%
temp/sources/eo/mutual{-}recursion{-}with{-}if{-}branching2{-}bad: test.derived: test.derived.m (was last redefined in "test.base") {-}> test.derived.n {-}> test.derived.m (was last redefined in "test.base")%
\item%
temp/sources/eo/mutual{-}recursion{-}with{-}if{-}branching2{-}good: test.derived: test.derived.m (was last redefined in "test.base") {-}> test.derived.n {-}> test.derived.m (was last redefined in "test.base")%
\item%
temp/sources/eo/mutual{-}recursion{-}with{-}if{-}branching3{-}bad: test.derived: test.derived.m (was last redefined in "test.base") {-}> test.derived.n {-}> test.derived.m (was last redefined in "test.base")%
\item%
temp/sources/eo/mutual{-}recursion{-}with{-}if{-}branching3{-}good: test.derived: test.derived.m (was last redefined in "test.base") {-}> test.derived.n {-}> test.derived.m (was last redefined in "test.base")%
\item%
temp/sources/eo/mutual{-}recursion{-}with{-}random{-}if{-}branching{-}bad: test.derived: test.derived.o (was last redefined in "test.base") {-}> test.derived.m {-}> test.derived.o (was last redefined in "test.base")%
\item%
temp/sources/eo/mutual{-}recursion{-}with{-}random{-}if{-}branching{-}good: test.derived: test.derived.o (was last redefined in "test.base") {-}> test.derived.m {-}> test.derived.o (was last redefined in "test.base")%
\end{enumerate}

\end{document}